\documentclass{article}
\usepackage{spconf,amsmath,graphicx}
\usepackage{graphicx}
\usepackage{caption}
\usepackage{subcaption}
\usepackage{algorithm}
\usepackage{algorithmic}

\usepackage{amsmath,amssymb}
\usepackage[keeplastbox]{flushend}
\usepackage{balance}
\usepackage[all=normal,paragraphs=tight,floats=normal,mathspacing=tight,
wordspacing=tight,charwidths=tight,mathdisplays=normal,
leading=normal]{savetrees}

\title{Deep learning for location based beamforming with NLOS channels}
\name{Luc Le Magoarou$^{\star }$ \qquad Taha Yassine$^{\star \dagger}$ \qquad St\'ephane Paquelet$^{\star }$ \qquad  Matthieu Crussi\`ere$^{\star \dagger}$\thanks{This work has been partly funded by the European Commission through the H2020 project Hexa-X (Grant Agreement no. 101015956).}}
  
  \address{$^{\star}$ b\raisebox{0.2mm}{\scalebox{0.7}{\textbf{$<>$}}}com, Rennes, France \\
      $^{\dagger}$Univ Rennes, INSA Rennes, IETR - UMR 6164 F-35000 Rennes, France}

\usepackage{algorithm}
\usepackage{algorithmic}

\usepackage{amsmath,amssymb}
\usepackage{graphicx}
\usepackage{caption}
\usepackage{subcaption}
\usepackage[keeplastbox]{flushend}

\usepackage{xcolor}
\newcommand{\new}[1]{\textcolor{black}{#1}}

\begin{document}
%
\title{Deep learning for location based beamforming with NLOS channels}
%
%
%

%
%

\markboth{}{}%

%



\maketitle

\begin{abstract}
Massive MIMO systems are highly efficient but critically rely on accurate channel state information (CSI) at the base station in order to determine appropriate precoders. CSI acquisition requires sending pilot symbols which induce an important overhead. 
In this paper, a method whose objective is to determine an appropriate precoder from the knowledge of the user’s location only is proposed. Such a way to determine precoders is known as location based beamforming. It allows to reduce or even eliminate the need for pilot symbols, depending on how the location is obtained. the proposed method learns a direct mapping from location to precoder in a supervised way. It involves a neural network with a specific structure based on random Fourier features allowing to learn functions containing high spatial frequencies. It is assessed empirically and yields promising results on realistic synthetic channels. As opposed to previously proposed methods, it allows to handle both line-of-sight (LOS) and non-line-of-sight (NLOS) channels.
\end{abstract}



%
\maketitle

\section{Introduction}
Machine learning (ML) techniques have been applied successfully to wireless communications in recent years (see \cite{Oshea2017,Wang2017} for exhaustive surveys). In particular, channel estimation and beamforming in the context of massive MIMO systems have benefited from ML \cite{Ding2018,Wei2019,Lemagoarou2020}. Although such approaches are very promising, their main drawback is that most need pilot symbols to be sent in order to estimate channel state information (CSI) or beam training procedures \cite{Barati2016}. Both approaches induce a consequent overhead and limit overall system efficiency.

Is CSI or beam training absolutely needed for a massive MIMO system to operate? Not necessarily, since appropriate precoders could be chosen at the base station based on other sources of information about users. In particular, it has been proposed to determine precoders based on estimated user locations, giving rise to the so-called location based beamforming (LBB) \cite{Maiberger2010,Kela2016,Abdelreheem2018}. For such approaches, the location of users is directly used to determine precoders. This allows to spare a lot of resources for communications and enhances the physical layer security \cite{Yan2015}. However, the main drawback of LBB with respect to CSI based approaches is that it assumes existence of a line of sight (LOS) propagation path on which the precoder is based. This LOS path may not exist in some scenarios, thus limiting the applicability of LBB. Nevertheless, the central element of LBB is the mapping that associates each location to a suitable precoder. This mapping can be learned by a neural network, in which case it is not limited to LOS channels.

\noindent {\bf Contributions.} In this paper, a supervised learning method allowing to map the user location to an appropriate precoder is proposed. It requires a labeled database containing user locations and the associated channels in order to train a neural network realizing the mapping. The obtained precoders handle with similar success LOS and NLOS channels. The neural network has a specific structure based on random Fourier features (RFF) \cite{Rahimi2007} that allows it to learn functions containing high spatial frequencies. This is beneficial to the task at hand because of the very fast variations of channel coefficients with respect to the user location due to fading.

Note that the location determination step is not studied in this paper, the location being considered known at the base station. This makes the proposed method totally independent of the specific location determination method that is used. In practice, location determination can be done either with help of a global navigation satellite system (GNSS), using a radar, with video cameras looking at the scene, or by sending pilot symbols (much less than for CSI acquisition) in order to determine directions of arrival.

\noindent {\bf Related work.} Location based beamforming has been proposed a decade ago \cite{Maiberger2010}. In its original formulation, it relies only on the main angle of departure (estimated by sending pilots) in order to synthesize an estimated single path channel which is then used to determine a precoder (without any learning). Such an approach is inherently limited to LOS channels, or at least channels comprising a dominant path. LBB Methods that were subsequently proposed \cite{Yan2015,Kela2016,Abdelreheem2018} introduce refinements to the original idea but do not get rid of this inherent limitation. The method proposed in this paper overcomes this limitation to LOS channels since it directly learns the location/precoder mapping, but as a counterpart, it requires a training phase to calibrate this mapping. {\color{black} The proposed approach pertains to integrated sensing and communication (ISAC) \cite{Liu2021}, and more specifically to sensing-assisted communication which aims at leveraging sensors at the base station to enhance communication \cite{Ali2020}.}

RFFs \cite{Rahimi2007} are a crucial element of the proposed neural network. Their usage in neural networks stems from the impossibility to learn mappings containing high frequencies with classical neural networks such as the multilayer perceptron (MLP) \cite{Tancik2020}. Coordinate based neural networks whose objective is to learn a mapping starting from a very low-dimensional space particularly benefit from RFFs \cite{Tancik2020}. RFF based neural networks have been applied successfully to 3D scene reconstruction \cite{Mildenhall2020}, where the network learns the mapping linking the 3D location and orientation to the radiance and density (which can be seen as the light channel).

\section{Problem formulation}

The method proposed in this paper applies to a wide variety of multi-user massive multiple input multiple output (massive MIMO) wideband systems \cite{Rusek2013,Larsson2014,Lu2014}, operating indifferently in time division duplex (TDD) or frequency division duplex (FDD), where the antennas at the base station are indifferently colocated or not (in which case it is a distributed MIMO system). Let us consider $A$ base station antennas and a single subcarrier, and denote $\mathbf{h} \in \mathbb{C}^{A}$ the downlink channel vector between the base station and any given user, and $\mathbf{l} \in \mathbb{R}^{D}$ its location, where $D$ can be two or three, depending on whether or not the elevation of users is considered relevant. 
Moreover, the proposed method can be directly transposed to systems comprising several subcarriers, but is presented here for a single subcarrier for ease of exposition.

Based on a labeled database of $N$ downlink channels associated with the corresponding user locations 
\begin{equation}
\left\{\mathbf{h}_i; \mathbf{l}_i\right\}_{i=1}^{N},
\label{eq:dataset}
\end{equation} 
the objective in this paper is to build a location based precoding function (or simply precoding function), whose function is to focus energy on the intended user location. It is mathematically defined as
\begin{equation}
\begin{array}{ccccl}
\mathcal{P} & : & \mathbb{R}^{D} & \to & \mathbb{C}^{A} \\
 & & \mathbf{l} & \mapsto & \mathbf{w} \triangleq\mathcal{P}(\mathbf{l}), \\
\end{array}
\label{eq:charting_function}
\end{equation}
where $\mathbf{w}$ is the predicted precoding vector (of unit norm).

\noindent{\bf Performance measure.}
In order to evaluate precoding functions, the normalized correlation between the precoder $\mathbf{w}$ and the channel $\mathbf{h}$ is used. It is expressed as
\begin{equation}
\eta \triangleq \frac{|\mathbf{w}^H\mathbf{h}|^2}{\left\Vert \mathbf{h} \right\Vert_2^2}.
\label{eq:corr}
\end{equation}
It is between zero and one (for a perfect precoder), and is tightly linked to the downlink channel capacity (considering a single user), whose expression is 
$$
\log(1+\eta.\text{SNR}_{\text{opt}}),
$$
for received signal of the form $y = \sqrt{P}\mathbf{w}^H\mathbf{h}s+n$ where $n\sim \mathcal{N}(0,\sigma^2)$ is additive noise, $s$ is the sent symbol and $P$ the transmit power. In that setting, $\text{SNR}_{\text{opt}} \triangleq \frac{P\left\Vert \mathbf{h} \right\Vert_2^2}{\sigma^2}$ is the highest achievable signal to noise ratio (SNR)  \cite{Bjornson2017}. In summary, the correlation is a single number between zero and one allowing to determine the maximum achievable downlink spectral efficiency for any transmit power and noise variance considering a given precoder.

\section{Proposed solution}
\label{sec:proposed_solution}

\noindent{\bf Neural precoding function.} Deep neural networks are known to be universal function approximators \cite{Cybenko1989,Hornik1989} and have led to great practical successes \cite{Lecun2015}. It is proposed here to implement the precoding function $\mathcal{P}$ as a deep neural network. However, it has been shown that classical neural networks known as multilayer perceptrons (MLP) \cite{Rumelhart1986} are in practice unable to learn functions of high frequency \cite{Tancik2020}. This phenomenon is known as spectral bias \cite{Rahaman2019,Basri2019} and has been characterized with help of the theory of neural tangent kernels (NTK) \cite{Jacot2018}.

\begin{figure}[tbp]
\centering
\includegraphics[width=0.5\columnwidth]{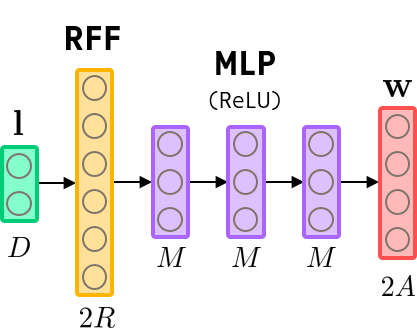}
\caption{Proposed RFF based neural network (with $Q=4$).}
\label{fig:network}
\end{figure}

In order to remedy this fundamental weakness, it has been proposed \cite{Tancik2020} to use random Fourier features \cite{Rahimi2007} to help the neural network building high frequencies. This line is followed in this paper, motivated by the fact that the optimal precoder may vary fast with respect to the user's location, due to fading. This results in a neural network whose first layer is fixed and corresponds to random Fourier features expressed as
$$
\gamma(\mathbf{l})=[\cos (2 \pi \mathbf{B} \mathbf{l}), \sin (2 \pi \mathbf{B} \mathbf{l})]^T,$$
where $\mathbf{B}\in \mathbb{R}^{R \times D}$ contains $R$ $D$-dimensional (spatial) frequencies drawn randomly as
$$  \mathbf{B} \sim \mathcal{N}(0,s^2 \mathbf{I}),$$
where the variance parameter $s^2$ controls the frequency range (the higher $s^2$ the more high frequencies are likely to appear).
It is proposed to concatenate this RFF layer with a classical MLP using fully connected layers and rectified linear units (ReLU) activation functions \cite{Nair2010} of width $M$ (number of neurons per layer) and depth $Q$ (number of layers). The last layer is of width $2A$, which corresponds to the real and imaginary parts of the predicted precoder $\mathbf{w}$ stacked together. The structure of the proposed neural network is shown on Fig.\ \ref{fig:network}. Simply put, the role of the RFF layer is to
pre-build high frequencies in order to ease learning for the subsequent MLP.

In order to calibrate the weights, the cost function to minimize is expressed as
\begin{equation}
\mathsf{CF} \triangleq 1-\frac{1}{N}\sum_{i=1}^N\frac{|\mathcal{P}(\mathbf{l}_i)^H\mathbf{h}_i|^2}{\left\Vert \mathbf{h}_i \right\Vert_2^2}.
\label{eq:cost}
\end{equation}
It measures the misalignment of the predicted precoders with respect to the training channels and is between $0$ (if the precoders are perfectly aligned with the channels) and $1$ (if the precoders are orthogonal to the channels). 

\noindent{\bf Practical setup.} Two distinct phases are to be distinguished for the proposed method to operate in a real system:
\begin{itemize}
\item \noindent{\bf Training phase:} At first, the database $\left\{\mathbf{h}_i; \mathbf{l}_i\right\}_{i=1}^{N}$ has to be built. To do so, the base station can send pilots and estimate channels using any classical method. Moreover, it has to access users' locations, using either a GNSS, video cameras or also pilots. The neural network can be trained with the built database in order to calibrate the precoding function $\mathcal{P}$.
\item \noindent{\bf Inference phase:} Then, once the neural network is properly trained, no more channel estimation is needed. The base stations only has to estimate users locations and is then able to predict appropriate precoders with the calibrated precoding function. 
\end{itemize}
Note that these two phases are not mutually exclusive, and can be intertwined in an online mode, in order to allow the base station to adapt to slow changes in the environment.

\section{Experiments}
\label{sec:experiments}
In this section, the proposed method is empirically assessed and compared to concurrent approaches on realistic multipath synthetic channels.

\noindent {\bf Simulation settings.} The experiments are carried out using training \new{multipath} channels taken from the DeepMIMO dataset \cite{Alkhateeb2019}. The `O1' urban outdoor ray-tracing scenario is chosen (with $5$ paths per channel, see \cite[Figure 2]{Alkhateeb2019} for a schematic view of the scenario), with a single base station (BS $16$) and a subset of all possible user locations (rows R$1000$ to R$3852$). The base station is equipped with a square uniform planar array (UPA) with $A=64$ half-wavelength separated antennas at a frequency of $3.5\,\text{GHz}$.

\noindent {\bf Implementation details.} The method is implemented with help of the PyTorch library \cite{Paszke2019}, so that gradients are computed automatically. The optimization of the precoding function is done according to the cost given in \eqref{eq:cost} on a database of $N=10000$   gradient descent (minibatches of $100$ training channels) using the Adam optimization algorithm \cite{Kingma2014} for $50$ epochs. The number of RFF is set to $R=1000$, and the standard deviation used for their computation is chosen by cross validation and fixed so that $\frac{1}{s}=50\,\text{m}$. The MLP following the RFF computation is of depth $Q=4$ and width $M=512$. 
Note that complex weights and inputs are handled classically by stacking the real and imaginary parts so that the neural network treats only real numbers.

\noindent {\bf Baselines.} The proposed method is compared with two baselines:
\begin{itemize}
\item A classical LBB approach \cite{Maiberger2010,Kela2016}, for which the precoder is chosen as the normalized LOS channel in the direction of the user, assuming the azimuth and elevation are perfectly known. This approach would be optimal in terms of correlation for LOS channels comprising a single path. This baseline is here to show the potential of the proposed method with respect to prior art.
\item A deep learning approach using a simple MLP (without RFFs). The architecture of the MLP is exactly the same as the one of the proposed method, except the first layer (RFF computation) that is replaced by a simple fully connected layer. This baseline is here to show the interest of RFFs.
\end{itemize}

\begin{figure}[t]
\includegraphics[width=\columnwidth]{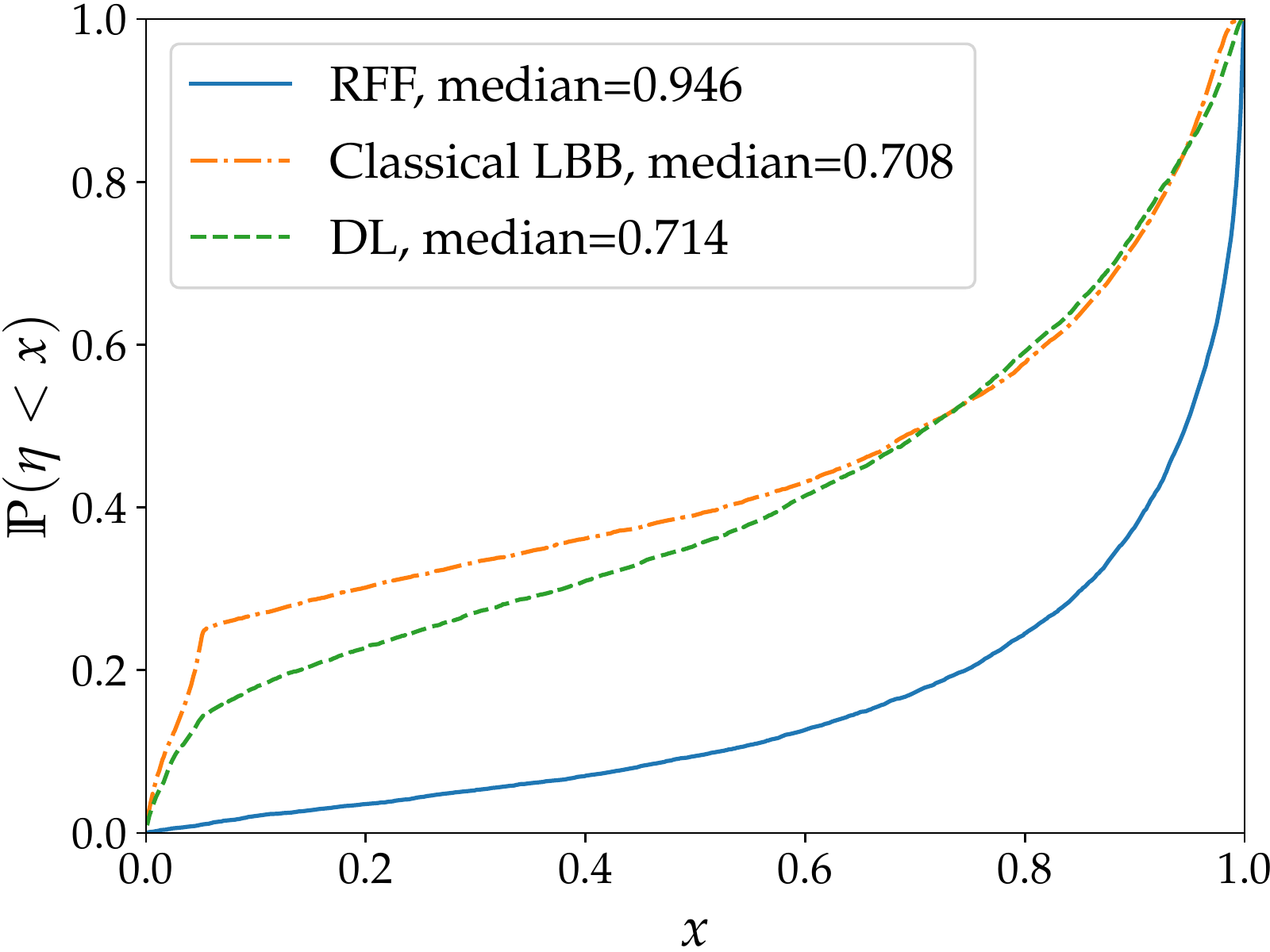}
\caption{CDF of the correlation between the channel and the precoder for several precoding methods.}
\label{fig:comp}
\end{figure}

\noindent {\bf Results.} The performance of the three compared methods on a test set of $10000$ channels (not used for training) is shown on Fig.\ \ref{fig:comp}. The cumulative distribution function (CDF) of the correlation  is plotted (with an ideal precoder it would be equal to zero for all values of $x<1$ and to one for $x=1$). The plot clearly shows the great improvement brought by the proposed method. Indeed, in such an urban environment where a LOS path is not available for every location, the classical LBB approach yields a median correlation of $0.708$. This is explained by the fact that the approach is inherently biased towards LOS channels. Using a deep learning approach with a simple neural network (without RFF) yields a slight improvement (median of $0.714$). This very simple neural network is prone to too much variance. However, the proposed method using RFFs allows to attain a median of $0.946$, which is a lot better than what is attained by the two aforementioned baselines, because of a better bias/variance trade-off.

\begin{figure}[t]
\includegraphics[width=\columnwidth]{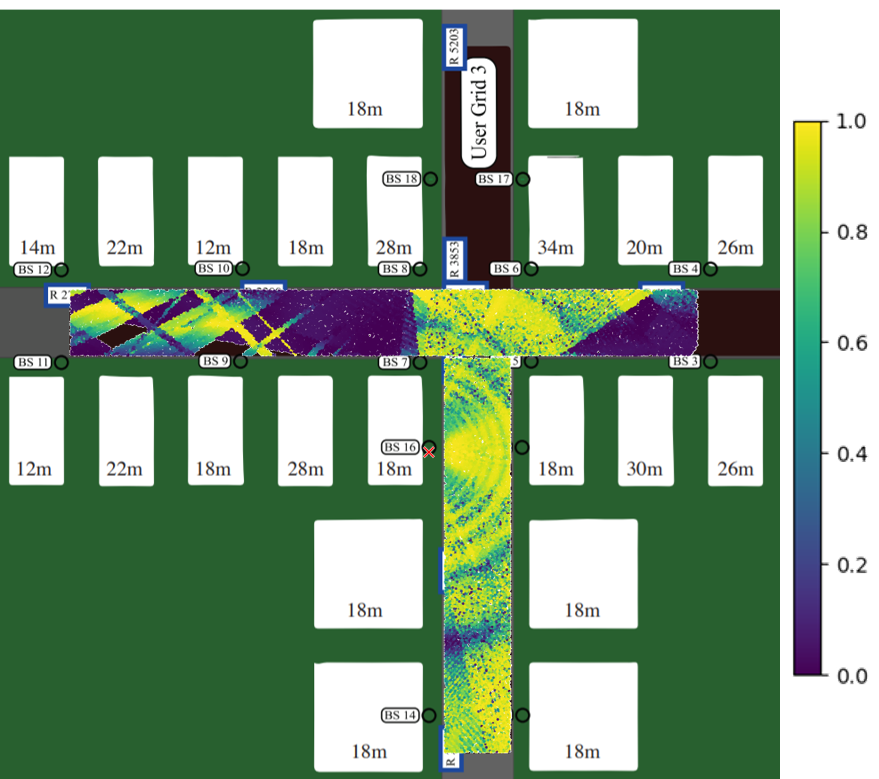}
\caption{Spatial performance of direction based LBB \cite{Maiberger2010,Kela2016}. Only LOS regions are well covered.}
\label{fig:res_LBB}
\end{figure}

\begin{figure}[t]
\includegraphics[width=\columnwidth]{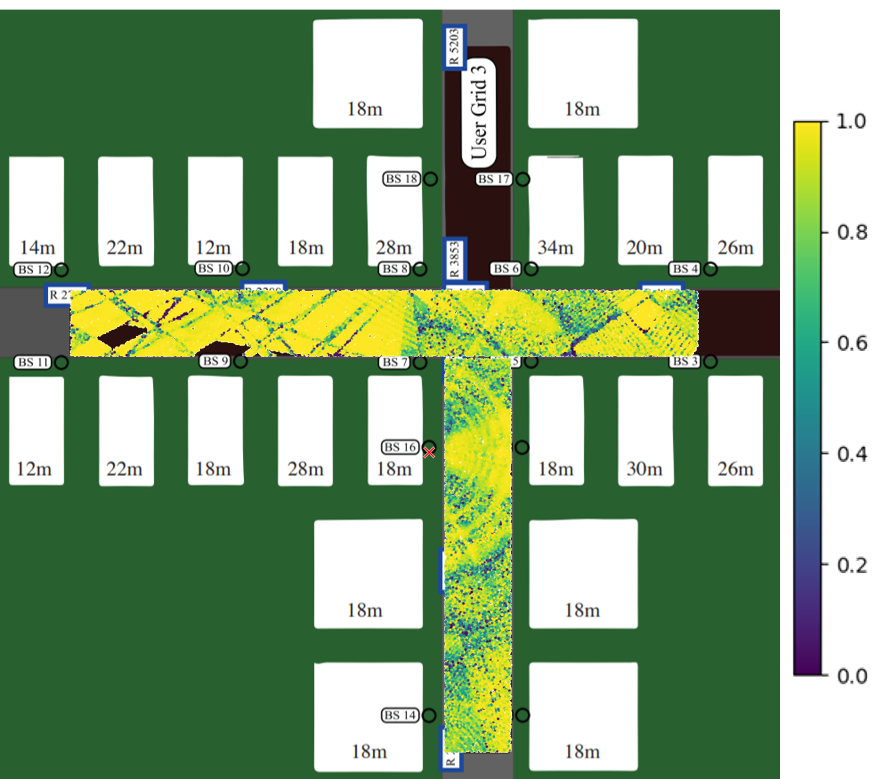}
\caption{Spatial performance of deep learning based LBB using RFFs (proposed). Both LOS and NLOS regions are well covered.}
\label{fig:res_RFF}
\end{figure}

It is also interesting to look at the spatial distribution of the attained correlation. This is depicted on Fig.\ \ref{fig:res_LBB} (for the classical LBB approach) and Fig.\ \ref{fig:res_RFF} (for the proposed approach). On these figures, the colors (blue to yellow) denote the obtained correlation for $50000$ test channels, and the base station location is denoted by a red cross. Note that on both figures, black areas on the left correspond to zero-norm channels. Comparing these figures, it is obvious that the classical LBB approaches (Fig.\ \ref{fig:res_LBB}) yield good precoders (yellow) for areas where user are in line of sight with respect to the base station, but very bad precoders elsewhere (blue areas). On the opposite, the proposed approach (Fig.\ \ref{fig:res_LBB}) yields good precoders everywhere on the considered area, even in locations that are in non-line of sight with respect to the base station. This clearly shows the benefit of the proposed method, which allows to obtain good precoders in areas where classical LBB approach fails, again owing to the incapacity of handling NLOS channels.

\begin{figure}[h!]
\centering
\includegraphics[width=\columnwidth]{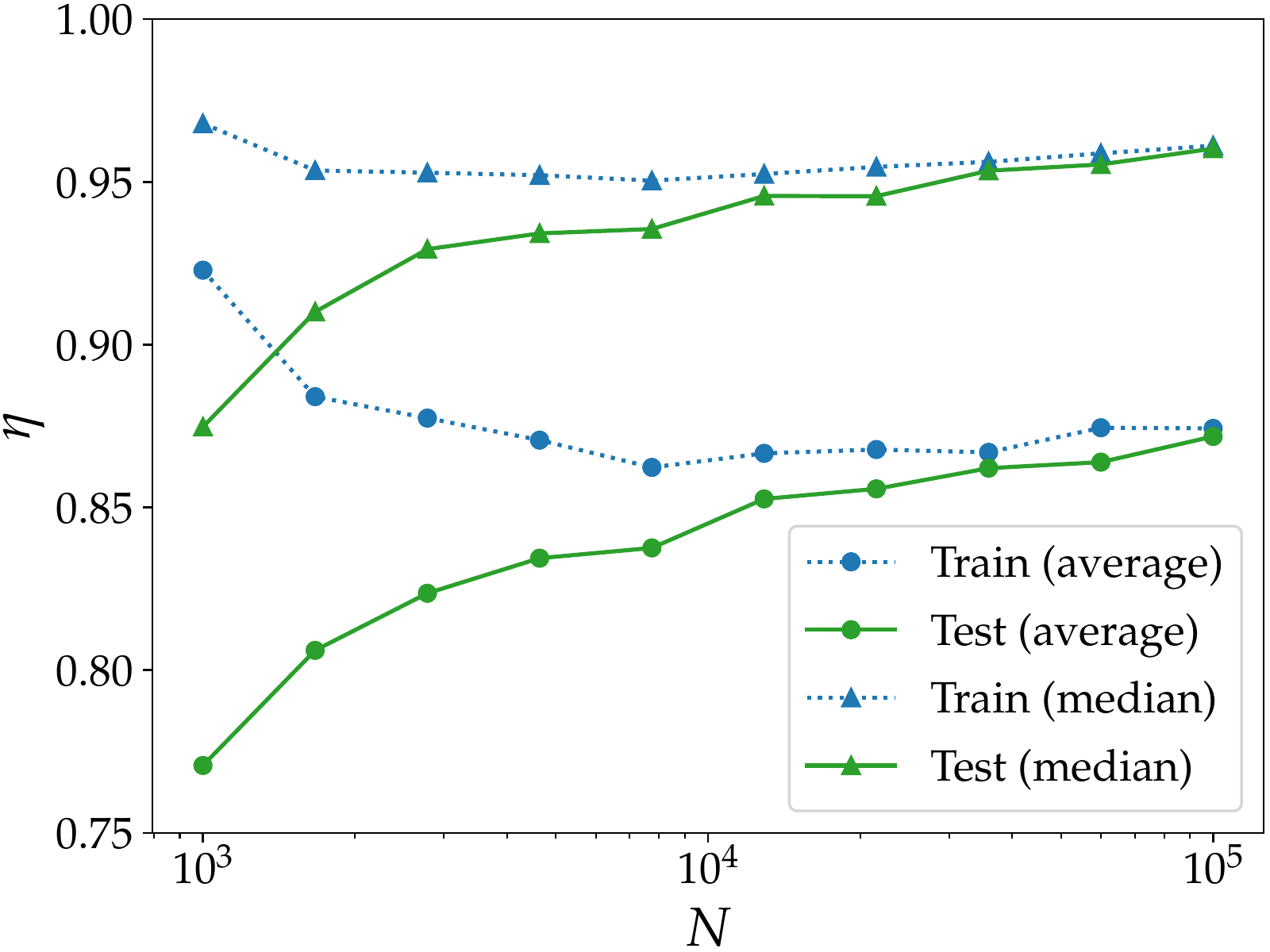}
\caption{Correlation versus number of training channels.}
\label{fig:N_train}
\end{figure}

 In order to assess the applicability of the proposed method, it is interesting to vary the number of training channels $N$ in order to determine which density of training channels is required. To that aim, the results in terms of average and median correlation are shown as a function of $N$ on Fig.\ \ref{fig:N_train}. From this figure, it can be seen that, obviously, the greater $N$ the greater the test correlation. However, one can see that the proposed method starts to perform well (on test data) for $N>3000$. The area in which users can be located spans a total surface of $20520\,\text{m}^2$, which means that the proposed method requires approximately one training channel every $\frac{20520}{3000} \approx 7\,\text{m}^2 $ on average to perform well. Such an empirical result is important in order to determine the time duration of the training phase in which both channels and locations have to be collected in order to build the database used to train the neural network.

\section{Conclusion}
\label{sec:conclusion}
In this paper, a location based beamforming method was proposed. It is based on a neural network comprising RFFs allowing to learn functions of high spatial frequency. The method was empirically assessed on realistic synthetic channels, yielding very promising results. Indeed, it is able to handle both LOS and NLOS channels whereas previously proposed methods perform well only for LOS channels. This shows the capacity of the proposed structure to interpolate well between training points thanks to RFFs which inject prior information on the learned mapping.

In the future, the method could be extended in several ways. First of all, the distribution of frequencies used to build the RFFs could be optimized to further enhance performance. It would also be interesting to aim at predicting not only an appropriate precoder but directly the channel vector. This would amount to summarize the whole propagation environment in the weights of a neural network, and could serve many applications, such as channel mapping \cite{Alrabeiah2019,Lemagoarou2021} (in space) or radio environment compression.



\bibliographystyle{unsrt}
{\small
\bibliography{biblio}}

\begin{thebibliography}{10}

\bibitem{Oshea2017}
Timothy O’Shea and Jakob Hoydis.
\newblock An introduction to deep learning for the physical layer.
\newblock {\em IEEE Transactions on Cognitive Communications and Networking},
  3(4):563--575, 2017.

\bibitem{Wang2017}
Tianqi Wang, Chao-Kai Wen, Hanqing Wang, Feifei Gao, Tao Jiang, and Shi Jin.
\newblock Deep learning for wireless physical layer: Opportunities and
  challenges.
\newblock {\em China Communications}, 14(11):92--111, 2017.

\bibitem{Ding2018}
Yacong Ding and Bhaskar~D Rao.
\newblock Dictionary learning-based sparse channel representation and
  estimation for fdd massive mimo systems.
\newblock {\em IEEE Transactions on Wireless Communications}, 17(8):5437--5451,
  2018.

\bibitem{Wei2019}
Xiuhong Wei, Chen Hu, and Linglong Dai.
\newblock Knowledge-aided deep learning for beamspace channel estimation in
  millimeter-wave massive mimo systems.
\newblock {\em arXiv preprint arXiv:1910.12455}, 2019.

\bibitem{Lemagoarou2020}
Luc {Le Magoarou} and Stéphane Paquelet.
\newblock Online unsupervised deep unfolding for massive mimo channel
  estimation, 2020.

\bibitem{Barati2016}
C.~Nicolas Barati, S.~Amir Hosseini, Marco Mezzavilla, Thanasis Korakis,
  Shivendra~S. Panwar, Sundeep Rangan, and Michele Zorzi.
\newblock Initial access in millimeter wave cellular systems.
\newblock {\em IEEE Transactions on Wireless Communications},
  15(12):7926--7940, 2016.

\bibitem{Maiberger2010}
Roy Maiberger, Doron Ezri, and Michael Erlihson.
\newblock Location based beamforming.
\newblock In {\em 2010 IEEE 26-th Convention of Electrical and Electronics
  Engineers in Israel}, pages 000184--000187. IEEE, 2010.

\bibitem{Kela2016}
Petteri Kela, Mario Costa, Jussi Turkka, Mike Koivisto, Janis Werner, Aki
  Hakkarainen, Mikko Valkama, Riku Jantti, and Kari Leppanen.
\newblock Location based beamforming in 5g ultra-dense networks.
\newblock In {\em 2016 IEEE 84th Vehicular Technology Conference (VTC-Fall)},
  pages 1--7. IEEE, 2016.

\bibitem{Abdelreheem2018}
Ahmed Abdelreheem, Ehab~Mahmoud Mohamed, and Hamada Esmaiel.
\newblock Location-based millimeter wave multi-level beamforming using
  compressive sensing.
\newblock {\em IEEE Communications Letters}, 22(1):185--188, 2018.

\bibitem{Yan2015}
Shihao Yan and Robert Malaney.
\newblock Location-based beamforming for enhancing secrecy in rician wiretap
  channels.
\newblock {\em IEEE Transactions on Wireless Communications}, 15(4):2780--2791,
  2015.

\bibitem{Rahimi2007}
Ali Rahimi and Benjamin Recht.
\newblock Random features for large-scale kernel machines.
\newblock In {\em Proceedings of the 20th International Conference on Neural
  Information Processing Systems}, NIPS'07, page 1177–1184, Red Hook, NY,
  USA, 2007. Curran Associates Inc.

\bibitem{Liu2021}
Fan Liu, Yuanhao Cui, Christos Masouros, Jie Xu, Tony~Xiao Han, Yonina~C.
  Eldar, and Stefano Buzzi.
\newblock Integrated sensing and communications: Towards dual-functional
  wireless networks for 6g and beyond.
\newblock {\em arXiv:2108.07165}, 2021.

\bibitem{Ali2020}
A.~Ali, N.~Gonzalez-Prelcic, R.W. Heath, and A.~Ghosh.
\newblock Leveraging sensing at the infrastructure for mmwave communication.
\newblock {\em IEEE Communications Magazine}, 58(7):84–89, 2020.

\bibitem{Tancik2020}
Matthew Tancik, Pratul~P Srinivasan, Ben Mildenhall, Sara Fridovich-Keil,
  Nithin Raghavan, Utkarsh Singhal, Ravi Ramamoorthi, Jonathan~T Barron, and
  Ren Ng.
\newblock Fourier features let networks learn high frequency functions in low
  dimensional domains.
\newblock {\em arXiv preprint arXiv:2006.10739}, 2020.

\bibitem{Mildenhall2020}
Ben Mildenhall, Pratul~P Srinivasan, Matthew Tancik, Jonathan~T Barron, Ravi
  Ramamoorthi, and Ren Ng.
\newblock Nerf: Representing scenes as neural radiance fields for view
  synthesis.
\newblock In {\em European Conference on Computer Vision}, pages 405--421.
  Springer, 2020.

\bibitem{Rusek2013}
Fredrik Rusek, Daniel Persson, Buon~Kiong Lau, Erik~G Larsson, Thomas~L
  Marzetta, Ove Edfors, and Fredrik Tufvesson.
\newblock Scaling up mimo: Opportunities and challenges with very large arrays.
\newblock {\em IEEE Signal Processing Magazine}, 30(1):40--60, 2013.

\bibitem{Larsson2014}
Erik~G Larsson, Ove Edfors, Fredrik Tufvesson, and Thomas~L Marzetta.
\newblock Massive mimo for next generation wireless systems.
\newblock {\em IEEE Communications Magazine}, 52(2):186--195, 2014.

\bibitem{Lu2014}
Lu~Lu, Geoffrey~Ye Li, A~Lee Swindlehurst, Alexei Ashikhmin, and Rui Zhang.
\newblock An overview of massive mimo: Benefits and challenges.
\newblock {\em IEEE journal of selected topics in signal processing},
  8(5):742--758, 2014.

\bibitem{Bjornson2017}
Emil Bj{\"o}rnson, Jakob Hoydis, Luca Sanguinetti, et~al.
\newblock Massive mimo networks: Spectral, energy, and hardware efficiency.
\newblock {\em Foundations and Trends{\textregistered} in Signal Processing},
  11(3-4):154--655, 2017.

\bibitem{Cybenko1989}
George Cybenko.
\newblock Approximation by superpositions of a sigmoidal function.
\newblock {\em Mathematics of control, signals and systems}, 2(4):303--314,
  1989.

\bibitem{Hornik1989}
Kurt Hornik, Maxwell Stinchcombe, and Halbert White.
\newblock Multilayer feedforward networks are universal approximators.
\newblock {\em Neural networks}, 2(5):359--366, 1989.

\bibitem{Lecun2015}
Yann LeCun, Yoshua Bengio, and Geoffrey Hinton.
\newblock Deep learning.
\newblock {\em nature}, 521(7553):436--444, 2015.

\bibitem{Rumelhart1986}
D.~E. Rumelhart, G.~E. Hinton, and R.~J. Williams.
\newblock {\em Learning Internal Representations by Error Propagation}, page
  318–362.
\newblock MIT Press, Cambridge, MA, USA, 1986.

\bibitem{Rahaman2019}
Nasim Rahaman, Aristide Baratin, Devansh Arpit, Felix Draxler, Min Lin, Fred
  Hamprecht, Yoshua Bengio, and Aaron Courville.
\newblock On the spectral bias of neural networks.
\newblock In {\em International Conference on Machine Learning}, pages
  5301--5310. PMLR, 2019.

\bibitem{Basri2019}
Ronen {Basri}, David {Jacobs}, Yoni {Kasten}, and Shira {Kritchman}.
\newblock {The Convergence Rate of Neural Networks for Learned Functions of
  Different Frequencies}.
\newblock {\em arXiv e-prints}, page arXiv:1906.00425, June 2019.

\bibitem{Jacot2018}
Arthur Jacot, Franck Gabriel, and Cl\'{e}ment Hongler.
\newblock Neural tangent kernel: Convergence and generalization in neural
  networks.
\newblock In {\em Proceedings of the 32nd International Conference on Neural
  Information Processing Systems}, NIPS'18, page 8580–8589, Red Hook, NY,
  USA, 2018. Curran Associates Inc.

\bibitem{Nair2010}
Vinod Nair and Geoffrey~E. Hinton.
\newblock Rectified linear units improve restricted boltzmann machines.
\newblock In {\em Proceedings of the 27th International Conference on
  International Conference on Machine Learning}, ICML'10, page 807–814,
  Madison, WI, USA, 2010. Omnipress.

\bibitem{Alkhateeb2019}
A.~Alkhateeb.
\newblock {DeepMIMO}: A generic deep learning dataset for millimeter wave and
  massive {MIMO} applications.
\newblock In {\em Proc. of Information Theory and Applications Workshop (ITA)},
  pages 1--8, San Diego, CA, Feb 2019.

\bibitem{Paszke2019}
Adam Paszke, Sam Gross, Francisco Massa, Adam Lerer, James Bradbury, Gregory
  Chanan, Trevor Killeen, Zeming Lin, Natalia Gimelshein, Luca Antiga, et~al.
\newblock Pytorch: An imperative style, high-performance deep learning library.
\newblock In {\em Advances in neural information processing systems}, pages
  8026--8037, 2019.

\bibitem{Kingma2014}
Diederik~P Kingma and Jimmy Ba.
\newblock Adam: A method for stochastic optimization.
\newblock {\em arXiv preprint arXiv:1412.6980}, 2014.

\bibitem{Alrabeiah2019}
Muhammad Alrabeiah and Ahmed Alkhateeb.
\newblock Deep learning for tdd and fdd massive mimo: Mapping channels in space
  and frequency.
\newblock In {\em 2019 53rd Asilomar Conference on Signals, Systems, and
  Computers}, pages 1465--1470. IEEE, 2019.

\bibitem{Lemagoarou2021}
L.~{Le Magoarou}.
\newblock Similarity-based prediction for channel mapping and user positioning.
\newblock {\em IEEE Communications Letters}, pages 1--1, 2021.

\end{thebibliography}
\end{document}